\documentclass{aa}  
\usepackage{graphicx}
\usepackage{txfonts}
\usepackage{natbib}
\bibpunct{(}{)}{;}{a}{}{,} 

\begin{document}

\title{Experimental oscillator strengths of \ion{Al}{i} lines for near-infrared astrophysical spectroscopy}
\author{M. Burheim\inst{1,}\inst{2}, H. Hartman\inst{2}, H. Nilsson\inst{2}}

   \institute{Lund Observatory, Division of Astrophysics, Department of Physics, Lund University, SE-221 00 Lund, Sweden \\
              \email{madeleine.burheim@fysik.lu.se; madeleine.burheim@mau.se}
         \and
             Department of Material Science and Applied Mathematics, Malm\"o University, 
                SE-205 06, Malm\"o, Sweden\\
             \email{henrik.hartman@mau.se, hampus.nilsson@mau.se}
             }
\titlerunning{Experimental oscillator strengths of \ion{Al}{i} lines for near-infrared astrophysical spectroscopy}
\authorrunning{M. Burheim et al.}
\date{Received XXX / Accepted YYY}
 \abstract
   {Elemental abundances can be determined from stellar spectra, making it possible to study galactic formation and evolution. Accurate atomic data is essential for the reliable interpretation and modeling of astrophysical spectra. In this work, we perform laboratory studies on neutral aluminium. This element is found, for example,  in young, massive stars and it is a key element for tracing ongoing nucleosynthesis throughout the Galaxy. The near-infrared (NIR) wavelength region is of particular importance, since extinction in this region is lower than for optical wavelengths. This makes the NIR wavelength region a better probe for highly obscured regions, such as those located close to the Galactic center.}
   {We investigate the spectrum of neutral aluminium with the aim to provide oscillator strengths ($f$-values) of improved accuracy for lines in the NIR and optical regions ($670 - 4200$ nm). }
   {Measurements of high-resolution spectra were performed using a Fourier transform spectrometer and a hollow cathode discharge lamp. The $f$-values were derived from experimental line intensities combined with published radiative lifetimes.}
   {We report oscillator strengths for 12 lines in the NIR and optical spectral regions, with an accuracy between 2 and 11\%, as well as branching fractions for an additional 16 lines.}
   {}

    \maketitle

\section{Introduction}

\begin{figure*}
\centering
   \includegraphics[width=20cm]{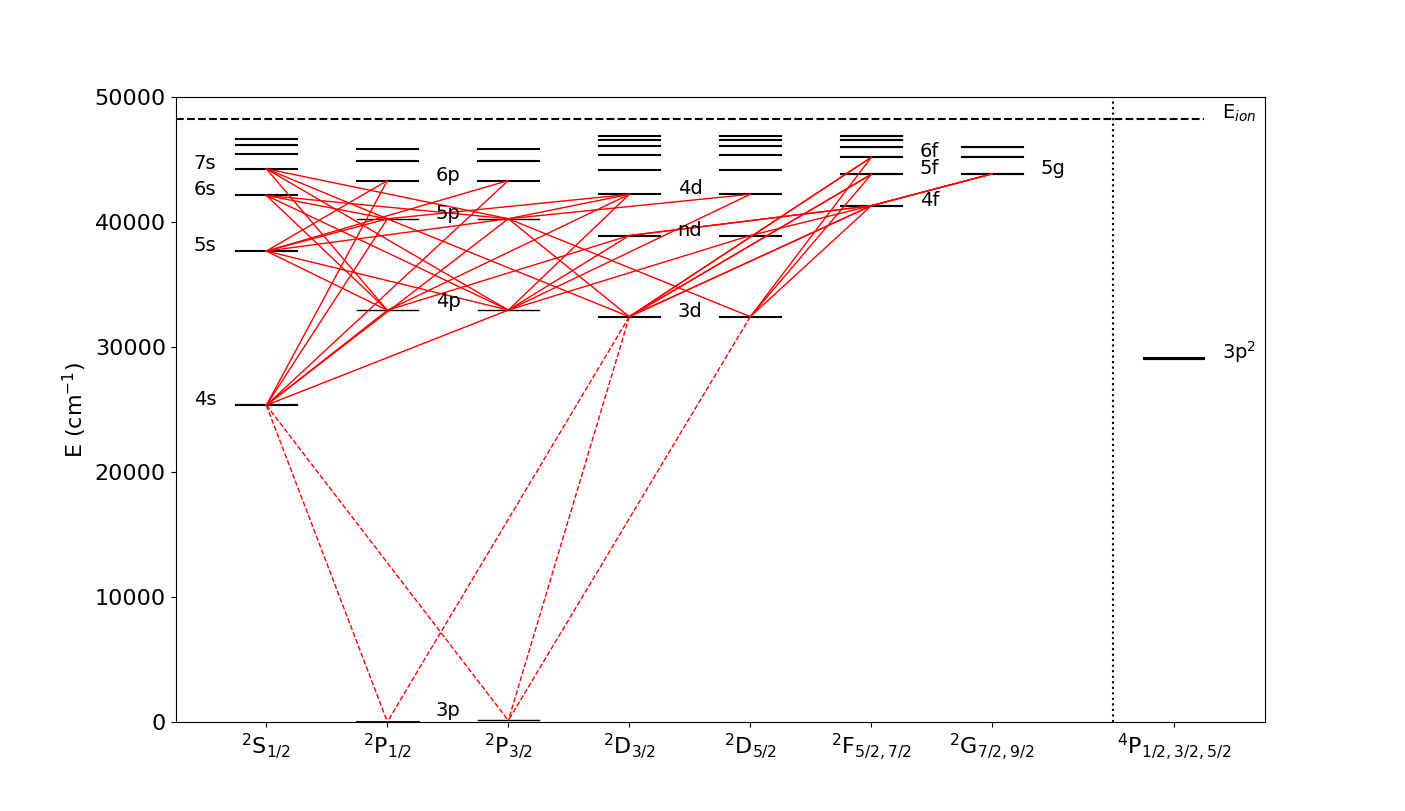}
\caption{Partial energy level diagram of \ion{Al}{i} with excitation energy (in cm$^{-1}$) on the vertical axis and the different terms in $LS$ notation along the horizontal axis. The red lines represent transitions observed in this project. The dashed red lines are transitions that were observed but not included in the analysis. Dashed black line at the top represents the ionisation energy.}
     \label{fig:energyleveldiagram}
\end{figure*}
From stellar evolution models, we know that stars create heavier elements in their core. At a later stage of their evolution, the stellar material is released into the interstellar medium. From this material, new stars are formed, which contain a higher abundance of heavy elements than the previous generation of stars. In this way, with each new generation, the amount of heavy elements in a galaxy increases. Thus, the chemical composition of a star provides information about the environment in which it was born. This makes it possible to understand the formation and evolution of our Galaxy.

The formation of the stable isotope $^{27}$Al requires an excess of neutrons, which can be provided in the core-collapse supernova or pair instability supernova of a massive star \citep{nordlander17}. The formation process of aluminium in different stellar populations is, however, still highly dubious and the key to fully understanding the physical processes behind its formation is accurate abundance analysis, which, in turn, depends on reliable atomic data. Aluminium yields seem to be strongly dependent on stellar mass and the ratio [Al/Fe], just as [$\alpha$/Fe], shows a clear distinction between the thin and thick Galactic disks \citep{bensby14}. 
Aluminium abundances in Globular Cluster (GC) stars can be used to understand their formation \citep{carretta18}. An enhancement in the aluminium abundance in metal-poor stars in the inner regions of the Galaxy shows a chemical signature similar to some second-generation stars, suggesting that they were formed in dissolved or evaporated GCs \citep{trincado20}. 
    
The infrared (IR) wavelength region is of particular importance, since the extinction is lower than for optical wavelengths. This makes the near-IR (NIR) wavelengths more suitable for studying regions close to the Galactic center \citep[e.g.,][]{thorsbro20,feldmeier17,do18}.
However, NIR lines are highly sensitive to stellar parameters, which means that they are more difficult to model (see e.g., \citet{nordlander17}). Therefore, accurate high-resolution atomic data play an important role in the reliable determination of abundances using such lines. 
        
Several studies on transition data for \ion{Al}{i} have been carried out in the past, focusing on the UV and optical spectral regions, and especially transitions involving the ground state. However, experimental data in the IR region is scarce. Relevant studies include lifetimes of the $^2$S$_{1/2}$ and $^2$D$_{3/2, 5/2}$ sequences measured by \citet{jonsson83} and the $^2$P$_{1/2, 3/2}$ sequence measured by \citet{jonsson84}, using laser induced fluorescence. \citet{buurman86} used laser excitation to determine the lifetimes and experimental absorption oscillator strengths of a number of levels and transitions in the lower part of the term system. In addition, they reported calculations using the code developed by \citet{cowan81}. The lifetimes of the 4p $^2$P levels were measured by \citet{buurman90} together with corresponding branching ratios. \citet{davidson90} measured lifetimes and derived oscillator strengths, using similar techniques, for levels along the $n$d $^2$D series. Furthermore, \citet{vujnovic02} reported lifetimes as well as transition rates for UV and optical \ion{Al}{i} lines.
Theoretical calculations have been reported by \citet{kurucz_database}, and more recently by \citet{papoulia19}. Both studies reported lifetimes and transition rates that are relevant to this work. 

Several studies \citep[e.g.,][]{nordlander17,heiter21,lind22} have identified aluminium lines that are useful for abundance determination in stellar spectra and pointed out the need for improved experimental oscillator strengths. 
We targeted some of these lines to be used directly in the chemical analysis, as well as additional lines as benchmarks for atomic structure calculations. 

In this work, we present $\log{gf}$-values and branching ratios for lines in the optical and IR spectral regions. The measurements were performed using a Fourier transform spectrometer (FTS) and the resulting branching fractions, together with known lifetimes from the literature, were used to derive transition probabilities. In Section 2, we describe the measurements together with the data analysis. The $\log{gf}$-values with uncertainties are presented in Section 3, together with comparisons with  data from previous studies.

\begin{figure}
\resizebox{\hsize}{!}{\includegraphics{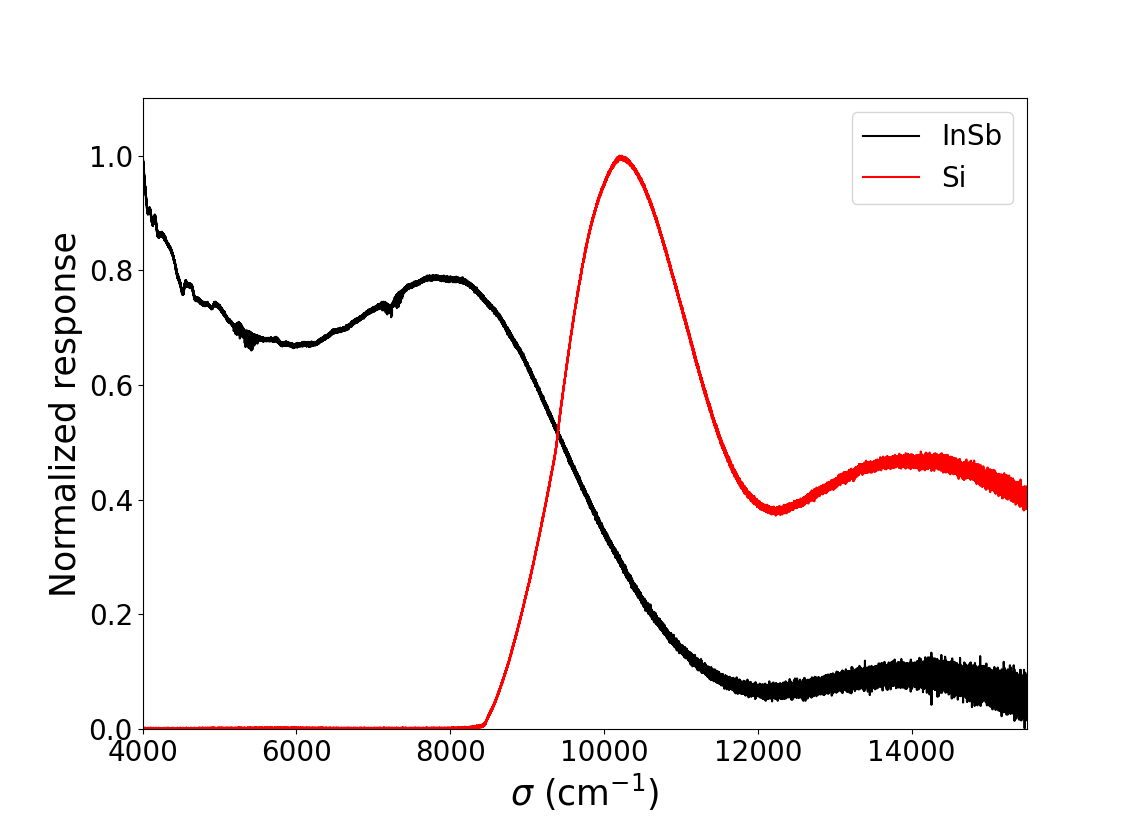}}
\resizebox{\hsize}{!}{\includegraphics{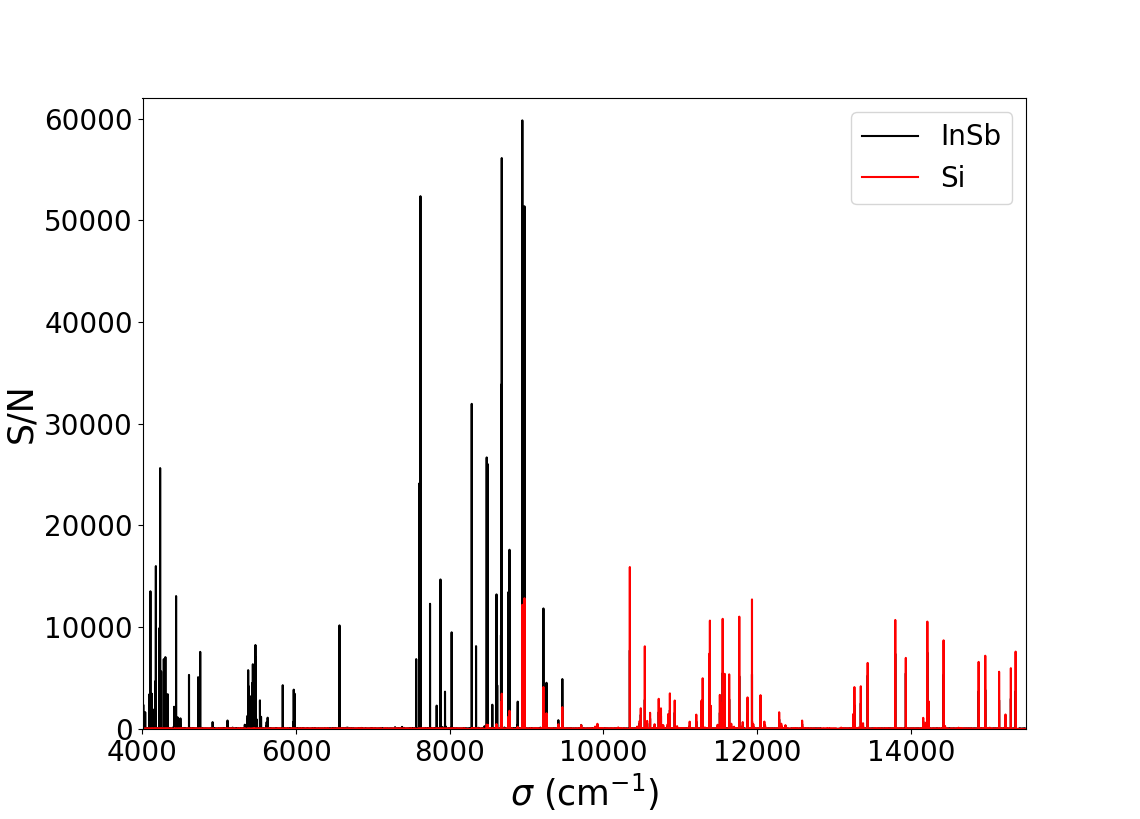}}
        \caption{Response curves (top) and spectra (bottom) for the two detector settings, using the InSb and Si detectors.} 
\label{fig:resp-InSb_Si}
\end{figure}

\begin{figure}
\resizebox{\hsize}{!}{\includegraphics{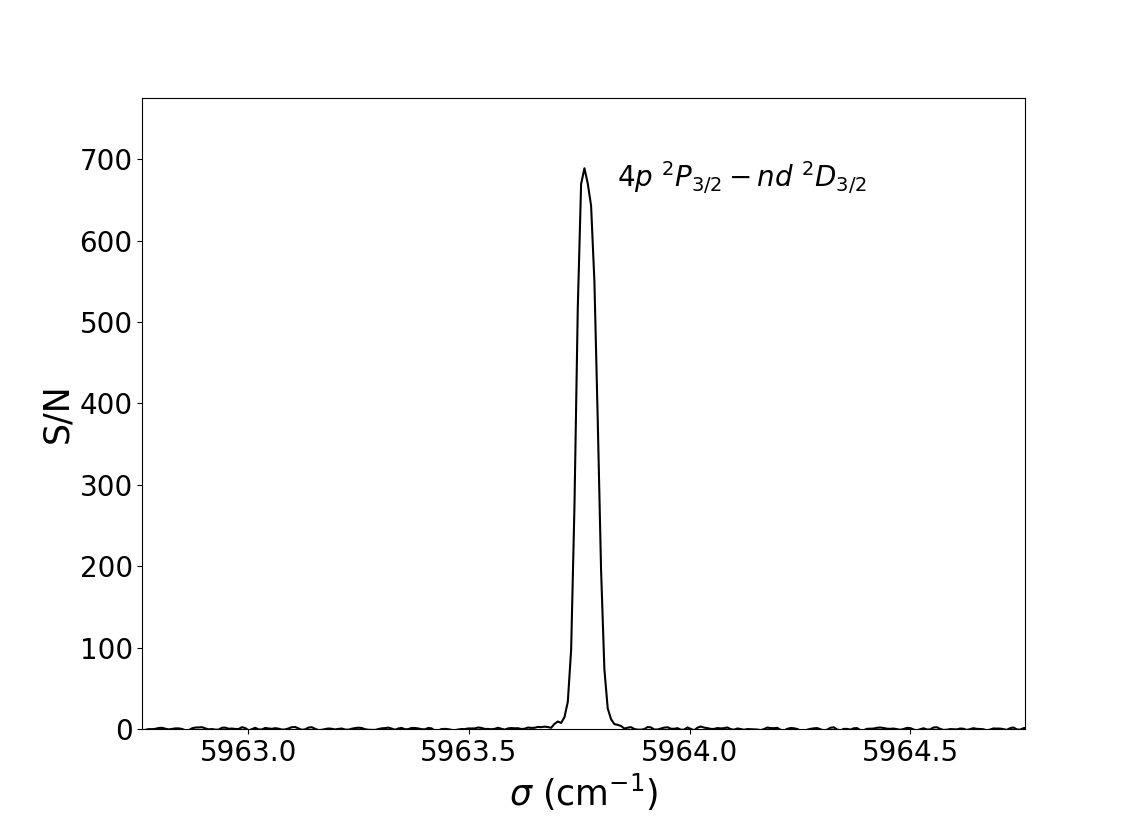}}
\resizebox{\hsize}{!}{\includegraphics{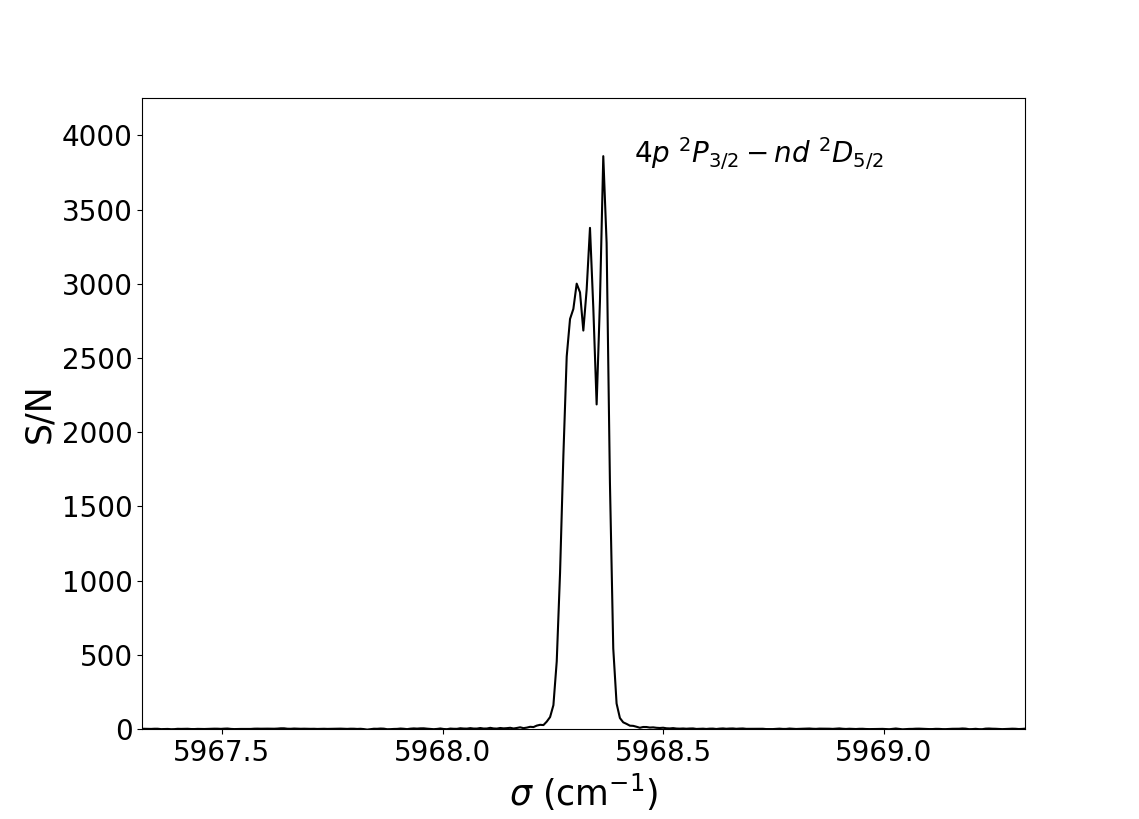}}
\caption{Lines from upper levels nd $^2$D$_{3/2}$ and nd $^2$D$_{5/2}$, respectively, where in the latter case we see resolved components of hyperfine structure.}
\label{fig:exampleLines}
\end{figure}
%
%
%
\section{Experimental oscillator strengths}
The atomic properties needed for quantitative analysis using spectral lines are (for absorption lines) the oscillator strength ($f$) and (for emission lines) the transition probability ($A$). These values can be derived from lifetimes and line intensity ratios, namely, the so-called branching fraction ($BF$).

The $BF$ for a line between levels $u$ and $l$ is defined as the transition probability ($A_{ul}$) of a transition divided by the sum of all transition probabilities ($A_{uj}$) for transitions from the same upper level, \textit{u}. For optically thin lines, $A_{ul}$ is proportional to the emitted intensity ($I_{ul}$), while the $BF$ for a transition is related to the relative intensity, such that:
\begin{equation}
BF_{ul} = \frac{A_{ul}}{\sum_{j}{A_{uj}}}=\frac{I_{ul}}{\sum_{j}{I_{uj}}}.
\label{eq:BF}
\end{equation}

The intensity, $I,$ can be measured from the recorded spectrum. The sum in Eq. \ref{eq:BF} includes all lines from the common upper level. However, in some cases not all lines can be measured,  either because they are too weak or  they lie in a spectral region not covered by the instrument. The total contribution of missing lines is called the residual and can be estimated using theoretical calculations. If the residual is large, reliable experimental $BF$s can not be derived, since the uncertainty in the residual will dominate the total uncertainty.

The $BF$, together with the radiative lifetime, $\tau_u = 1/\sum_k{A_{uk}}$, of the upper level, gives the transition probability:
\begin{equation}
A_{ul}=\frac{BF_{ul}}{\tau_u}
\label{eq:A}
,\end{equation}
from which the oscillator strength ($f$-value) can be derived as:
\begin{equation}
f = 1.499\cdot 10^{-14}\frac{g_u}{g_l}\mathbf{\lambda_{vac}}^2A_{ul},
\label{eq:fvalue}
\end{equation}
where $g_u$ and $g_l$ are the statistical weights of the upper and lower levels, respectively, the vacuum wavelength, $\mathbf{\lambda_{vac}}$, is in units of nm and $A_{ul}$ in s$^{-1}$ \citep{spectrophysics}.

\subsection{Measurements of branching fractions}
The experimental setup consists of two main components: a hollow cathode discharge lamp (HCL) used as light source and a spectrometer to record the emitted spectrum. The HCL consists of two anodes and a cylinder-shaped cathode, isolated by glass cylinders. The cathode used in this experiment was made of solid aluminium with an inner diameter of $7$ mm. On either side of the glass tube, there is an anode at a $20$ mm distance from the cathode. The spectra were recorded using the Lund Observatory high-resolution FTS, a Bruker IFS 125 HR, which covers a spectral range of 2000$-$50000 cm$^{-1}$ (5000$-$200 nm) and has a maximum resolving power of $\sigma/\Delta\sigma = 10^6$ at 2000 cm$^{-1}$. 

The aluminium spectra were produced using neon as carrier gas, at a pressure of 1 Torr, and currents ranging from 0.20 A to 0.70 A. 
The resolution of the Fourier transform spectrometer (FTS) was set to 0.02 cm$^{-1}$. The spectra were recorded using two detectors: an indium antimonide (InSb) detector and a silicon (Si) detector, sensitive in different spectral regions. This allows for a larger spectral coverage. The response in the different detector settings is shown in Fig. \ref{fig:resp-InSb_Si}, along with the recorded spectra.
 
The line intensities were determined as the integrated area of a spectral line and (where possible) the area of a Gaussian fit, using GFit \citep{gfit}. The former method was needed where unresolved hyperfine structure was present. For single-component lines, both techniques were used and consistent results were obtained. Examples of lines with and without hyperfine structure are shown in Fig. \ref{fig:exampleLines}.

\subsection{Intensity calibration}
The transmission of the spectrometer and the optical elements in the optical path is dependent on wavelength. This contributes to a difference in line intensity ratios between the observed and the intrinsic values and is referred to as the response function. In order to determine the response function, a tungsten filament lamp (Osram Wi 17/G) with known relative spectral intensity distribution was recorded with the same optical path as the emission line spectra. The tungsten lamp is calibrated for spectral radiance by the Swedish National Laboratory in the spectral range 4000$-$40000 cm$^{-1}$ (2500$-$250 nm). Comparing the measured lamp spectrum to the calibrated one, the response function of the instrument and detectors was derived (see Fig. \ref{fig:resp-InSb_Si}). The HCL and the tungsten lamp were placed at the same distance from the FTS and a folding mirror was used to change from one light source to the other. The tungsten lamp spectra were recorded immediately before and after each aluminium spectrum, in order to assure there had been no change in the instrument response during the measurements. The response function for each detector was used to calibrate the measured aluminium lines. Spectra from the two detectors were connected by putting their relative intensities on the same scale using \ion{Al}{i} lines in the overlapping region. The intensity calibration was applied to the measured line intensities.
In addition, to be able to calibrate lines below 4000 cm$^{-1}$, a black-body radiator with a temperature of 1200 $^\circ$C (Optronic Laboratories, OL480) was used. 

\begin{figure}
  \resizebox{\hsize}{!}{\includegraphics{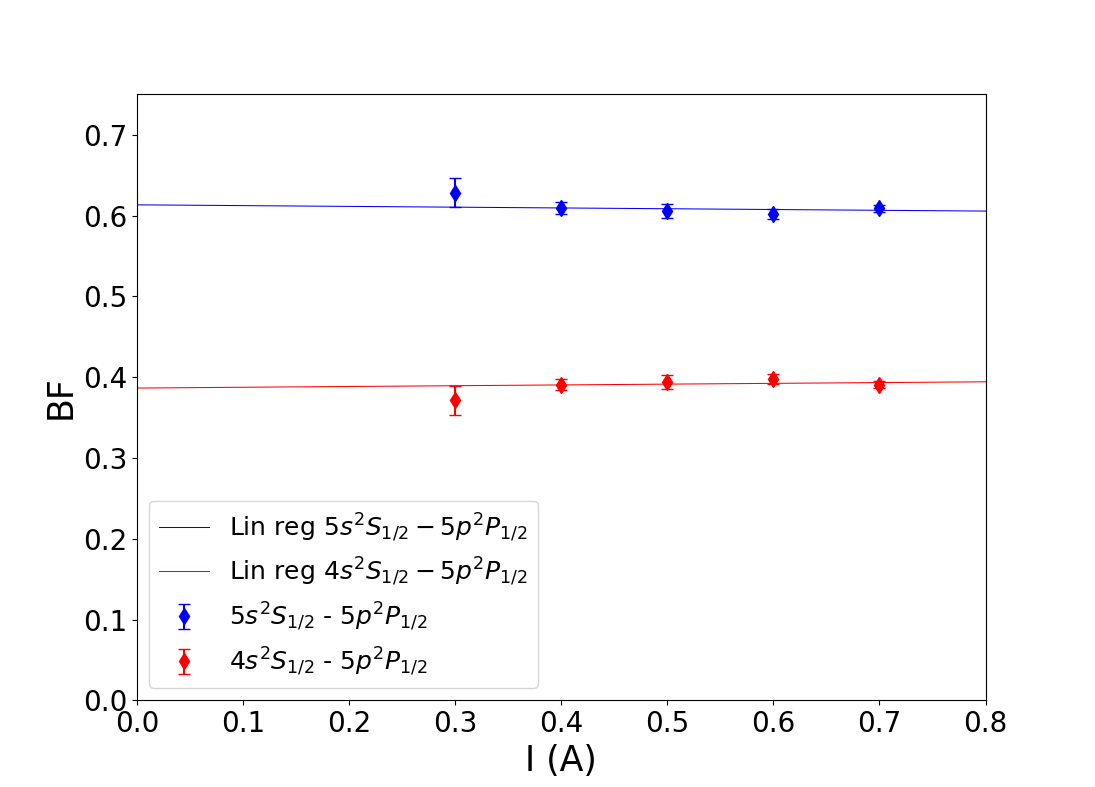}}
  \resizebox{\hsize}{!}{\includegraphics{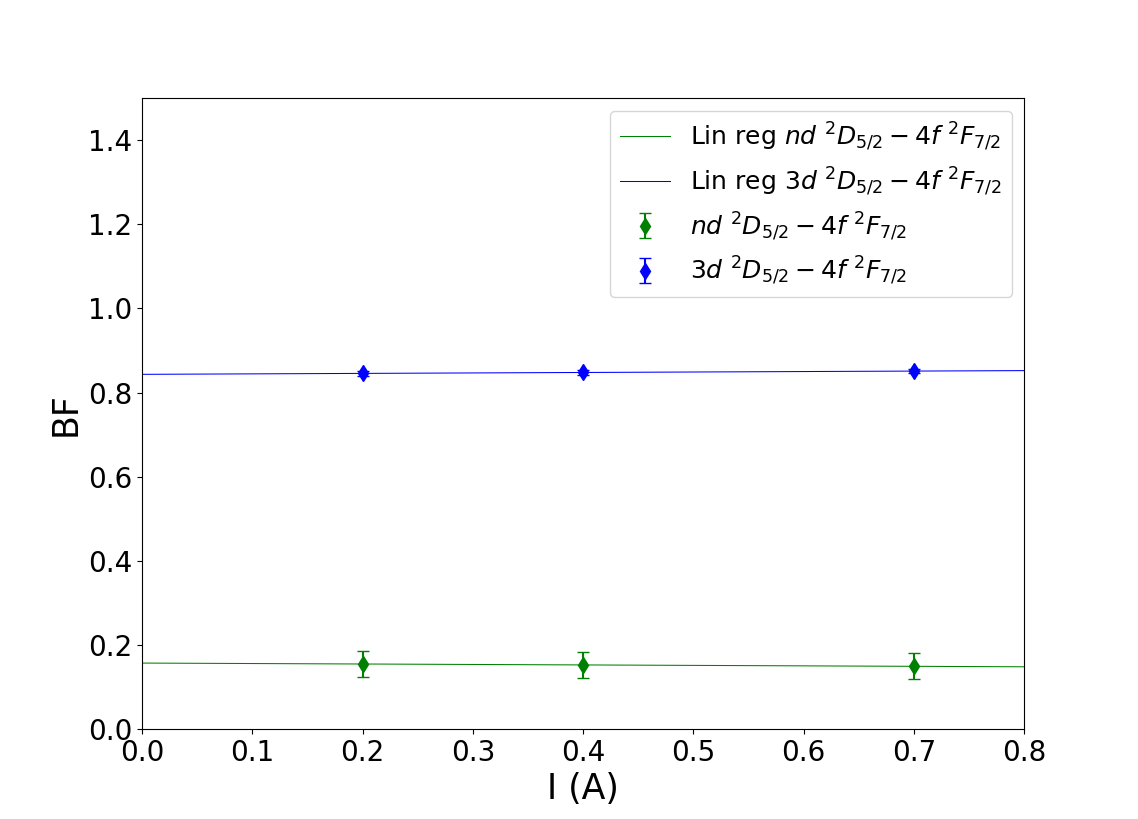}}
        \caption{Self-absorption analysis for transitions from upper levels 5p $^2$P$_{1/2}$ (top) and 4f $^2$F$_{7/2}$ (bottom), where the measured $BF$s are investigated as a function of current. A linear regression is fitted to the data points.} 
        \label{fig:selfabsorptionexamples}
\end{figure}
\subsection{Self-absorption}
An increased current through the HCL increases the amount of atoms and ions in the plasma, which leads to stronger emission lines. This is advantageous for the study of weak lines. However, for strong lines, an increased current may lead to self-absorption, that is: if the atom (or ion) density in the plasma is high enough, photons emitted in a transition to a lower level that is long-lived and highly populated may be reabsorbed. Self-absorption thus affects the observed line in that it shows an intensity that is too low, which affects the $BF$ measurements (see Eq. \ref{eq:BF}). The line profile is altered due to the difference in optical depth between the core and the wings.

Since the amount of self-absorption is dependant on the population in the lower level, self-absorption can be investigated by studying the ratio of the line intensities between two lines from the same upper level, as a function of current. If no self-absorption is present, the ratio should be constant with respect to current. If the ratio is not constant, it can be extrapolated to zero current, which corresponds to a self-absorption free value. The technique of interpolation is discussed in \citet{sikstrom02}. The uncertainty in the self-absorption can be estimated from the statistical uncertainty of the y-axis intercept for stronger lines and from the scatter in $BF$ for weaker lines.

Self-absorption was investigated for all levels included in this work. No self-absorption was detected for lines from the levels 5s, nd, 5p, and 4f. Examples are presented in Fig. \ref{fig:selfabsorptionexamples}. Transitions from the levels 6s, 4d, and 7s were too weak to be measured in spectra with low currents, preventing a full study of these as a function of current. However, since these transitions have the same lower levels as the transitions from 5s and nd, which are shown to be self-absorption free, we can conclude that there is no self-absorption present in the lines from the upper levels 6s, 4d, and 7s. 

Transitions to the ground configuration 3p are not included in our analysis, since these are strongly self-absorbed and even show a self-reversed line structure. Lines involving the 4s and nd levels show similar structure, but this is due to hyperfine splitting, which arises from the interaction with the nucleus (see Fig. \ref{fig:exampleLines}).

\subsection{Uncertainties}
As discussed in \citet{sikstrom02}, the uncertainty in the transition probabilities depends on the uncertainty in the branching fractions and the uncertainty in the radiative lifetime. Assuming that the uncertainties on the line intensity and the lifetime are uncorrelated, the uncertainty in the transition probability, and thus the oscillator strength, can be estimated as:
\begin{equation}
\left(\frac{u(A_k)}{A_k}\right)^2 = \left(\frac{u(f_k)}{f_k}\right)^2 = \left(\frac{u(BF_k)}{BF_k}\right)^2 + \left(\frac{u(\tau)}{\tau}\right)^2,
\end{equation}
where $u(\tau)$ is the uncertainty of the lifetime of the upper level. The uncertainty in the branching fraction of a line, $k$, measured with a detector, $P,$ is given by:
\begin{equation}
\begin{aligned}
\left(\frac{u(BF_k)}{BF_k}\right)^2 &=\  (1-BF_k)^2\left(\frac{u(I_k)}{I_k}\right)^2 \\ &+ \sum_{j\neq k (in P)}^{n}{(BF_j)^2}\left(\left(\frac{u(I_j)}{I_j}\right)^2 + \left(\frac{u(c_j)}{c_j}\right)^2\right) \\ &+ \sum_{j\neq k (in Q)}^{n}{(BF_j)^2}\left(\left(\frac{u(I_j)}{I_j}\right)^2 + \left(\frac{u(c_j)}{c_j}\right)^2 + \left(\frac{u(nf)}{nf}\right)^2\right).
\end{aligned}
\end{equation}
The uncertainty in the branching fraction is a result of the uncertainty in line intensity measurement of the line itself, $u(I_k)$, and the uncertainty in the intensity measurements of all other lines from the same upper level, $u(I_j)$. The uncertainty in the line intensity measurement includes the uncertainty in the determination of the line area and the uncertainty due to self-absorption (see Section 2.4). The uncertainty from the intensity calibration, $u(c_j)$, adds to the uncertainty for all other lines. In the case where lines are measured with an additional detector, $Q$, an additional uncertainty, $u(nf)$, will arise due to the merging of the different spectra.

\section{Results and conclusions} 

Oscillator strengths for 12 lines and branching fractions for 16 lines are reported in Tables \ref{tab:BFtable} and \ref{tab:BRtable}, respectively. The data is presented in two sets, depending on the size of the residual $BF$s. For levels where the residual is small ($\leq10\%$), $BF$s were derived and, together with radiative lifetimes from the literature, transition probabilities, $A_{ul}$, and $\log{gf}$-values (see Table \ref{tab:BFtable}). In the case where the residual is larger than 10\%, $BF$s were derived but no $\log{gf}$s are reported. The uncertainty in these $BF$s were derived similarly to that of the first set of $BF$s, but without taking the uncertainty in the residual into account (see Table \ref{tab:BRtable}). The latter set of $BF$s cannot be directly used for stellar abundances, but they are still valuable for benchmarking theoretical calculations. 

\begin{figure}[h!]
\includegraphics[width=0.5\textwidth]{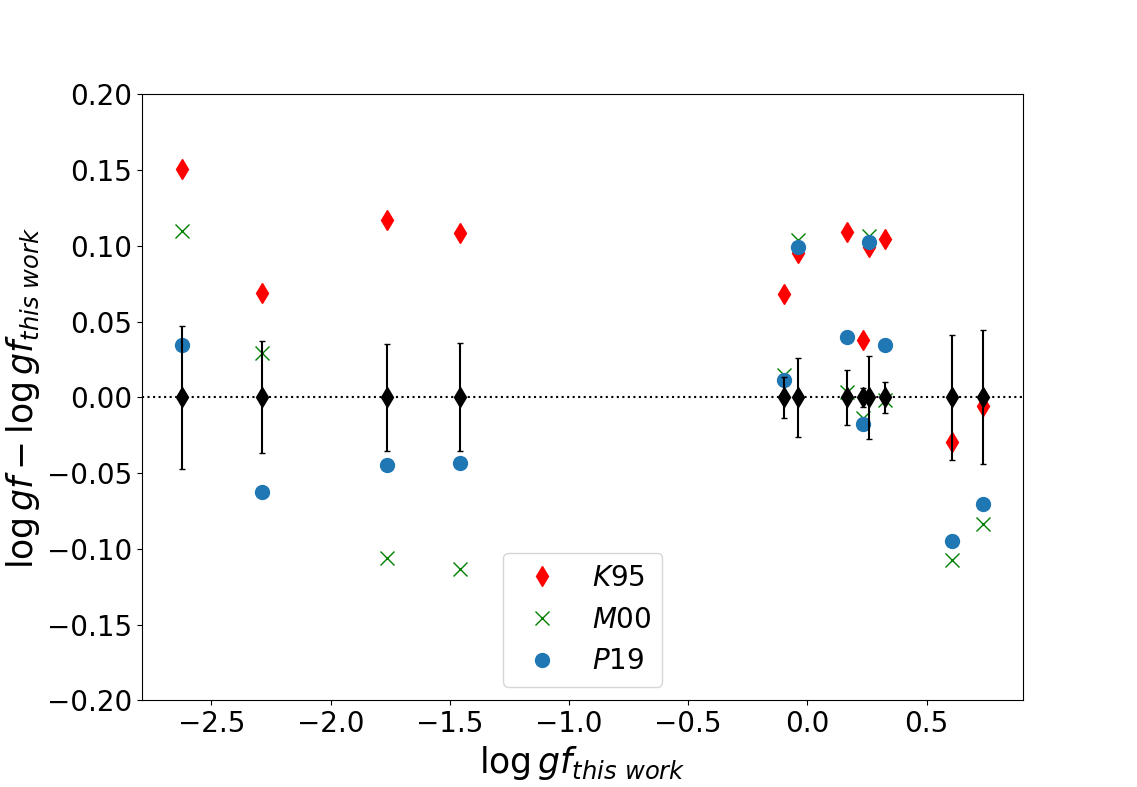}
\includegraphics[width=0.5\textwidth]{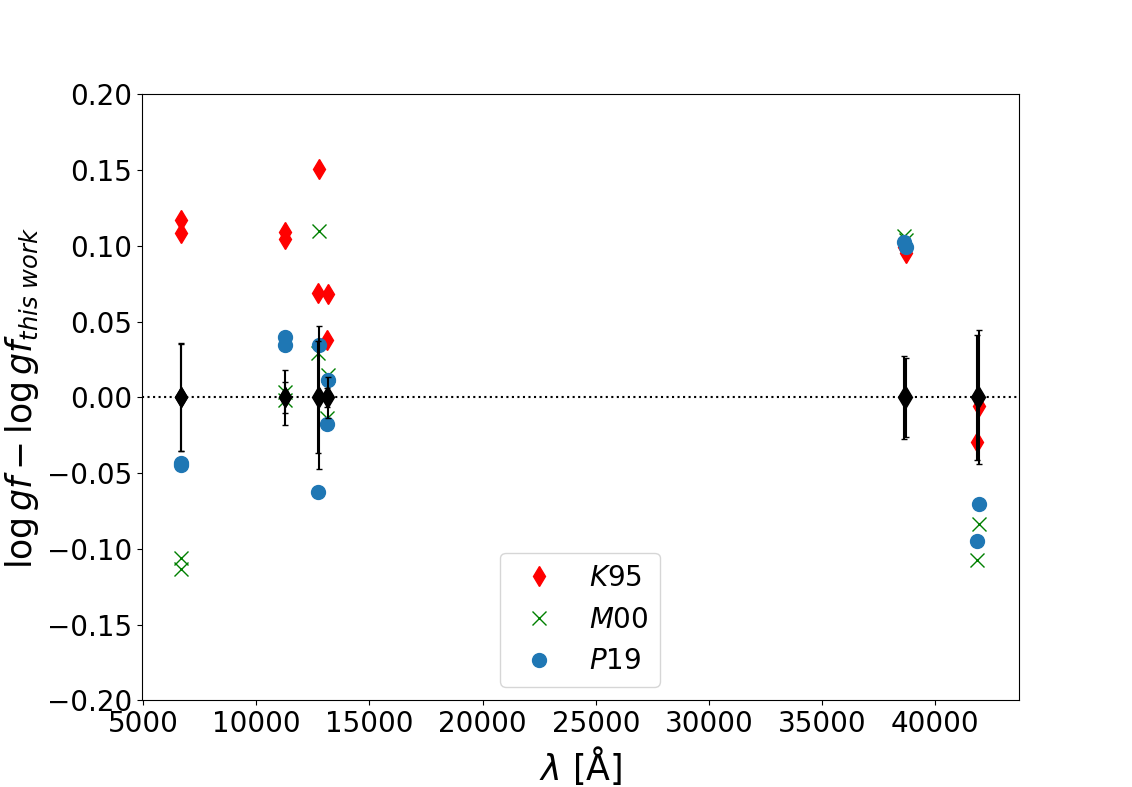}
    \caption{Comparison between the $\log{gf}$s derived in this work and values from the literature (\citet{kurucz_database}, \citet{topbase}, \citet{papoulia19}). The error bars represent our uncertainties. \label{fig:comparisonstheory}}
    \label{fig:exptheorcomp}
\end{figure}

The weaker lines (see Table \ref{tab:BFtable}) generally have larger uncertainties of $8-10\%$, whereas the stronger lines typically have smaller uncertainties of $<5\%$. The uncertainty contribution from the self-absorption analysis, for the transitions presented in Table \ref{tab:BFtable}, is  $2\%  $ on average. The uncertainty on the experimental lifetimes ranges between $3\%$ and $9\%,$ whereas for the theoretical lifetimes, we used d$T$ for the uncertainty, as given in \citet{papoulia19}. Table \ref{tab:loggf_comp} and Fig.\ \ref{fig:exptheorcomp} show a comparison between our experimental $\log{gf}$-values and theoretical values by \citet{kurucz_database}, \citet{topbase}, and \citet{papoulia19}. The difference between our experimental $\log{gf}$-values and those calculated by \citet{papoulia19} is, on average, $-0.001$ dex, with a standard deviation of $0.06$ dex. However, the deviation is larger for the calculations by \citet{kurucz_database} and \citet{topbase}. We see a smaller deviation from the theoretical values for the stronger lines, as well as those from the lower levels of excitation. Transitions including the nd levels show a larger deviation as compared to the calculations by \citet{papoulia19}, in spite of their higher transition probabilities\footnote{We chose to use the naming convention of the second term in the series as 3s$^2$nd $^2$D, noting that a large contribution is from the 3s3p$^2$ $^2$D term, which is nonetheless not its largest component. For further discussion, see \citet{papoulia19}.}.  

To our knowledge, no previous experimental oscillator strengths for the lines we present have been reported. The exception being the 4s$-$4p transitions, which we present in Table \ref{tab:BFtable}, where $f$-values were derived by \citet{buurman86}, however, these values are determined by the lifetime, since there is only one decay channel. 

\onecolumn
\newpage

\begin{table}
\caption{\label{tab:BFtable}Transition data for \ion{Al}{i}, for lines with derived $\log{gf}$ values. Lifetimes used in the analysis include experimental$^{\textup{b,c}}$ and theoretical$^{\textup{d}}$ sources. Residuals are derived from the calculated transition rates by \citet{papoulia19}.}
\centering
\begin{tabular}{cccclcccc}
\hline \hline
Upper             & Lower & $\sigma^{\textup{a}}$  &    $\lambda_{vac}^{\textup{a}}$  & $BF$$_{\textup{exp}}$  & Unc$_{BF}$ & A$_{ul}$ & $\log{gf}$$_{\textup{exp}}$ & Unc$_{gf}$ \\
level &     level &  [cm$^{-1}$]        &  [\AA]         &              &[\%]     &   [s$^{-1}$]  &                               & [\%] \\
\hline
4p $^2$P$_{1/2}$ & 4s $^2$S$_{1/2}$   & 7602.048  & 13154.350  & 0.9998 & 0.01 & $1.54\cdot 10^{7}$ & -0.0980 & 3.1 \\
                 & $Residual$          &                      &                   & 0.0002  &  &  &  \\ 
$\tau = 65(2)$ ns $^{\textup{b}}$ &   &               &                   &  &  &  &  \\ \hline 
                 &                        &                   &                   &  &  &  &  \\ 
4p $^2$P$_{3/2}$ & 4s $^2$S$_{1/2}$   & 7617.884  & 13127.005  & 0.9998 & 0.01 & $1.65\cdot 10^{7}$ & 0.232 & 1.5 \\
                 & $Residual$         &               &                   & 0.0002  &  &  &  \\
$\tau = 60.5(9)$ ns $^{\textup{b}}$ &  &                      &                   &  &  &  &  \\ \hline 
                 &                        &                   &                   &  &  &  &  \\ 
5p $^2$P$_{1/2}$ & 5s $^2$S$_{1/2}$   & 2582.565  &     38721.19 & 0.558 & 5 & $2.03\cdot 10^{6}$ & -0.0396 & 6 \\                      
& 3d $^2$D$_{3/2}$   & 7836.521  & 12760.765  &  0.013 & 10 & $4.88\cdot 10^{4}$ & -2.62 & 11 \\
                 & 4s $^2$S$_{1/2}$   & 14924.209 &     6700.522  & 0.352 & 8 & $1.28\cdot 10^{6}$ & -1.76 & 8 \\ 
                 & $Residual$         &               &                   & 0.077 &  &  \\
$\tau = 275(8)$ ns $^{\textup{c}}$ &  &               &                   &  &  &  &  \\ \hline 
                 &                        &                   &                   &  &  &  &  \\ 
5p $^2$P$_{3/2}$ & 5s $^2$S$_{1/2}$   & 2588.472  &     38632.83 & 0.557 & 6 & $2.03\cdot 10^{6}$ & 0.259 & 7 \\ 
                 & 3d $^2$D$_{5/2}$   & 7841.048  & 12753.397  & 0.015 & 8 & $5.31\cdot 10^{4}$ & -2.29 & 9 \\
                 & 4s $^2$S$_{1/2}$   & 14930.134 &     6697.864   & 0.358 & 8 & $1.30\cdot 10^{6}$ & -1.46 & 9 \\ 
                 & $Residual$         &               &                   & 0.078 &  &  &  &  \\
$\tau = 275(8)$ ns $^{\textup{c}}$ &  &               &                   &  &  &  &  \\ \hline 
                 &                        &                   &                   &  &  &  &  \\ 
4f $^2$F$_{5/2}$ & nd $^2$D$_{3/2}$   & 2389.976  &     41841.42 & 0.155 & 9 & $2.57\cdot 10^{6}$ & 0.607 & 10 $$*$$ \\ 
                 & 3d $^2$D$_{3/2}$   & 8883.937  &     11256.270  &  0.778 & 4 & $1.29\cdot 10^{7}$ & 0.167 & 5 $$*$$ \\ 
                 & $Residual$         &               &                   & 0.067 &  &  &  &  \\
$\tau = 60.4(12)$ ns $^{\textup{d}}$ &  &                     &                   &  &  &  &  &  \\ \hline  
                 &                        &                   &                   &  &  &  &  &  \\ 
4f $^2$F$_{7/2}$ & nd $^2$D$_{5/2}$   & 2385.46  &      41920.7 & 0.157$^{\textup{e}}$ & 10 & $2.59\cdot 10^{6}$ & 0.737 & 11 $$*$$ \\ 
                 & 3d $^2$D$_{5/2}$   & 8882.566  &     11258.008  &  0.843 & 1 & $1.40\cdot 10^{7}$ & 0.327 & 2 $$*$$ \\ 
                 & $Residual$         &               &                   & 0.000 &  &  &  \\
$\tau = 60.4(12)$ ns $^{\textup{d}}$ &  &                     &                   &  &  &  &  \\ \hline  
\hline   
\end{tabular}                                                                                         
\tablebib{$^{\textup{a}}$Wavenumbers and wavelengths are from the NIST database \citep{NIST_ASD}, based on studies by \citet{eriksson63}, \citet{biemont87_alt}, and \citet{chang90}. 
All wavelengths given are in vacuum, $\lambda_{vac}$.
$^{\textup{b}}$\cite{buurman90}; $^{\textup{c}}$\cite{buurman86}; $^{\textup{d}}$\cite{papoulia19}; $*$Using the uncertainty in the lifetime $u(\tau)$ from \citet{papoulia19}. }
\tablefoot{$^{\textup{e}}$This is blended by nd $^2$D$_{5/2}$$-$4f $^2$F$_{5/2}$, but the latter is predicted to be more than an order of magnitude weaker and is thus omitted. This is reflected in the uncertainty of this line. }
\end{table}

\newpage
\begin{table}
\caption{\label{tab:BRtable}Transition data for \ion{Al}{i}. Lifetimes used in the analysis include experimental$^{\textup{b,d}}$ and theoretical$^{\textup{c,e}}$ sources. Residuals have been derived from the calculated transition rates by \citet{papoulia19}, for all levels except 7s, for which the values from \citet{kurucz_database} were used. The residuals have not been taken into account when estimating Unc$_{BF}$.} 
\centering
\begin{tabular}{cccclcc}
\hline\hline
Upper             & Lower & $\sigma^{\textup{a}}$  &    $\lambda_{vac}^{\textup{a}}$  &       $BF$$_{\textup{exp}}$ & $BF$$_{\textup{theor}}$ & Unc$_{BF}$ \\
level &     level &  [cm$^{-1}$]        &  [\AA]        &       &   &[\%] \\
\hline
5s $^2$S$_{1/2}$ & 4p $^2$P$_{3/2}$   & 4723.759  & 21169.58 & 0.153 & 0.153 & 2 \\
                 & 4p $^2$P$_{1/2}$   & 4739.597  & 21098.84 & 0.077 & 0.077 & 5 \\
                 & $Residual$         &               &                    & & 0.770  &  \\
$\tau = 19.8(5)$ ns $^{\textup{b}}$ &  &                      &                     &  &  \\ \hline  
                 &                                &                       &                          &   \\
nd $^2$D$_{3/2}$ & 4p $^2$P$_{3/2}$   & 5963.759  &     16767.948  &     0.065 & 0.060 & 6 \\ 
                 & 4p $^2$P$_{1/2}$   & 5979.595  &     16723.541  &     0.295 & 0.301 & 1 \\ 
                 & $Residual$         &               &                    & & 0.639 & \\
$\tau = 29.5(7)$ ns $^{\textup{b}}$ &  &                      &                     &  \\ \hline  
                 &                        &                   &                     &  \\ 
6s $^2$S$_{1/2}$ & 5p $^2$P$_{3/2}$   & 1866.530  & 53575.35 & 0.064 & 0.087 & 7 \\
                 & 5p $^2$P$_{1/2}$   & 1872.437  & 53406.34 & 0.029 & 0.044 & 13 \\
                 & 4p $^2$P$_{3/2}$   & 9178.761  & 10894.716  &  0.141 & 0.117 & 4 \\
                 & 4p $^2$P$_{1/2}$   & 9194.596  & 10875.953  &  0.072 & 0.059 & 6 \\
                 & $Residual$         &               &                    & & 0.693 & \\
$\tau = 48.12(72)$ ns $^{\textup{c}}$ &  &                    &                     & \\ \hline  
                 &                                &                       &                          & \\
4d $^2$D$_{3/2}$ & 5p $^2$P$_{3/2}$   & 1955.851  &     51128.64 & 0.003 & 0.004 & 39 \\ 
                 & 5p $^2$P$_{1/2}$   & 1961.756  &     50974.74 & 0.024 & 0.021 & 5 \\ 
                 & 4p $^2$P$_{1/2}$   & 9283.919  &     10771.313  &     0.003 & 0.005 & 11 \\ 
                 & $Residual$         &               &                    & & 0.970 & \\ 
$\tau = 13.2(3)$ ns $^{\textup{d}}$ &  &                      &                     & \\ \hline  
                 &                        &                   &                     & \\ 
4d $^2$D$_{5/2}$ & 5p $^2$P$_{3/2}$   & 1959.909  &     51022.78 & 0.028 & 0.025 & 3 \\ 
                 & 4p $^2$P$_{3/2}$   & 9272.138  &     10785.000  &     0.003 & 0.006 & 31 \\ 
                 & $Residual$         &               &                    & & 0.969 & \\
$\tau = 13.2(3)$ ns $^{\textup{d}}$ &  &                      &                     & \\ \hline  
                 &                        &                   &                     & \\ 
7s $^2$S$_{1/2}$ & 5p $^2$P$_{3/2}$   & 3995.255  & 25029.69 & 0.046 & 0.046 $^{\textup{(e)}}$ & 11 \\
                 & 5p $^2$P$_{1/2}$   & 4001.160  & 24992.75 & 0.026 & 0.023 $^{\textup{(e)}}$ & 14 \\
                 & 4p $^2$P$_{3/2}$   & 11307.478 & 8843.705 &  0.077  & 0.079 $^{\textup{(e)}}$ & 7 \\
                 & $Residual$         &               &                    & & 0.851 & \\
$\tau = 72.27$ ns $^{\textup{e}}$ &  &                &                     & \\ \hline  
\hline                                                                                         
\end{tabular}
\tablefoot{We chose to use the naming convention of the second term in the series as 3s$^2$nd $^2$D, noting that a large contribution is from the 3s3p$^2$ $^2$D term, however not its largest component. For further discussion, see \citet{papoulia19}.}
\tablebib{$^{\textup{a}}$Wavenumbers and wavelengths are from the NIST database \citep{NIST_ASD}, based on studies by \citet{eriksson63}, \citet{biemont87_alt}, and \citet{chang90}. 
All wavelengths given are in a vacuum, $\lambda_{vac}$.
$^{\textup{b}}$\cite{buurman86}; $^{\textup{c}}$\cite{papoulia19}; $^{\textup{d}}$\cite{davidson90}; $^{\textup{e}}$\cite{kurucz_database}.}
\end{table}

\newpage

\begin{table}
\caption{\label{tab:loggf_comp}Comparison of our data to the theoretical $\log{gf}$-values from the literature.}
\centering
\begin{tabular}{cccccccc}
\hline\hline
Lower & Upper & $\sigma^{\textup{a}}$  &        $\lambda_{vac}^{\textup{a}}$  & $\log{gf}$$_{\textup{exp}}$ & $\log{gf}$$_{\textup{P19}}$ & $\log{gf}$$_{\textup{K95}}$ & $\log{gf}$$_{\textup{M00}}$ \\
level &     level &  [cm$^{-1}$]        &  [\AA]         &              &     &  & \\
\hline

4s $^2$S$_{1/2}$ & 5p $^2$P$_{3/2}$ & 14930.134 &       6697.864  & -1.46 & -1.499 & -1.347 & -1.569 \\
4s $^2$S$_{1/2}$ & 5p $^2$P$_{1/2}$ & 14924.209 &       6700.522  & -1.76 & -1.808 & -1.647 & -1.870 \\
3d $^2$D$_{3/2}$ & 4f $^2$F$_{5/2}$ & 8883.937  &       11256.270 & 0.167 & 0.206 & 0.276 & 0.170 \\ 
3d $^2$D$_{5/2}$ & 4f $^2$F$_{7/2}$ & 8882.566  &       11258.008 & 0.327 & 0.362 & 0.431 & 0.325 \\ 
3d $^2$D$_{5/2}$ & 5p $^2$P$_{3/2}$ & 7841.048  & 12753.397 & -2.29 & -2.348 & -2.217 & -2.257 \\
3d $^2$D$_{3/2}$ & 5p $^2$P$_{1/2}$ & 7836.521  & 12760.765 & -2.62 & -2.588 & -2.472 & -2.513 \\
4s $^2$S$_{1/2}$ & 4p $^2$P$_{3/2}$ & 7617.884  & 13127.005 & 0.232 & 0.215 & 0.270 & 0.219 \\
4s $^2$S$_{1/2}$ & 4p $^2$P$_{1/2}$ & 7602.048  & 13154.350 & -0.0980 & -0.0867 & -0.030 & -0.083 \\
5s $^2$S$_{1/2}$ & 5p $^2$P$_{3/2}$ & 2588.472  &       38632.83  & 0.259 & 0.361 & 0.358 & 0.365 \\ 
5s $^2$S$_{1/2}$ & 5p $^2$P$_{1/2}$ & 2582.565  &       38721.19  & -0.0396 & 0.0596 & 0.056 & 0.064 \\ 
nd $^2$D$_{3/2}$ & 4f $^2$F$_{5/2}$ & 2389.976  &       41841.42 & 0.607 & 0.513 & 0.578 & 0.500 \\ 
nd $^2$D$_{5/2}$ & 4f $^2$F$_{7/2}$ & 2385.46   &       41920.7 & 0.737 & 0.667 & 0.732 & 0.654 \\ 
                 \hline  
\hline   
\end{tabular}
\tablefoot{$^{\textup{a}}$Wavenumbers and wavelengths are from the NIST database \citep{NIST_ASD}, based on studies by \citet{eriksson63}, \citet{biemont87_alt}, and \citet{chang90}.
All wavelengths given are in a vacuum, $\lambda_{vac}$.}
\end{table}

\twocolumn

\begin{acknowledgements}
We acknowledge support from the Swedish Research Council VR through grant no 2016-04185, and the Crafoord foundation. The infrared FTS at the Edl{\'e}n laboratory was bought through a grant from the Knut and Alice Wallenberg Foundation. We also thank the referee for their constructive input which has helped improve the manuscript. 
\end{acknowledgements}

\bibliographystyle{aa} 
\bibliography{bibliography} 

\end{document}